\documentclass[sigconf]{acmart}

\usepackage{booktabs} 
\graphicspath{ {images/} }
\usepackage{caption}
\usepackage{subcaption}
\usepackage{hyperref}
\setcopyright{rightsretained}
\begin{document}
\title{Indian Regional Movie Dataset for Recommender Systems}
\author{Prerna Agarwal}\authornote{Prerna Agarwal (now working with IBM Research) and Richa Verma (now working with TCS Research ) have equal contribution}
\affiliation{
  \institution{IIIT Delhi}}
\email{prerna15045@iiitd.ac.in}
\author{Richa Verma\footnotemark[1]}
\affiliation{
  \institution{IIIT Delhi}}
\email{richa15054@iiitd.ac.in}
\author{Angshul Majumdar}
\affiliation{
  \institution{IIIT Delhi}}
\email{angshul@iiitd.ac.in}
\vspace{10 mm}
\begin{abstract}
Indian regional movie dataset is the first database of regional Indian movies, users and their ratings. It consists of movies belonging to 18 different Indian regional languages and metadata of users with varying demographics. Through this dataset, the diversity of Indian regional cinema and its huge viewership is captured. We analyze the dataset that contains roughly 10K ratings of 919 users and 2,851 movies using some supervised and unsupervised collaborative filtering techniques like Probabilistic Matrix Factorization, Matrix Completion, Blind Compressed Sensing etc. The dataset consists of metadata information of users like age, occupation, home state and known languages. It also consists of metadata of movies like genre, language, release year and cast. India has a wide base of viewers which is evident by the large number of movies released every year and the huge box-office revenue. This dataset can be used for designing recommendation systems for Indian users and regional movies, which do not, yet, exist. The dataset can be downloaded from \href{https://goo.gl/EmTPv6}{https://goo.gl/EmTPv6}.
\end{abstract}

%
%
%


\maketitle
\section{Introduction}
Recommendation systems [17, 18, 19, 20, 33] utilize user ratings to provide personalized suggestions of items like movies and products. Some popular brands that provide such services are Amazon [34], Netflix, IMDb, BarnesAndNoble [35], etc. Collaborative Filtering (CF) and Content-based (CB) recommendation are two commonly used techniques for building recommendation systems. CF systems [21, 22, 23]operate by gathering user ratings for different items in a given domain and compare various users, their similarities and differences, to determine the items to be recommended. Content-based methods recommend items by comparing representations of content that a user is interested in to representations of content that an item consists of. \\ 
There are several datasets like MovieLens~\cite{term1} and Netflix~\cite{term2} that are available for testing and bench-marking recommendation systems. \\
We present an Indian regional movie dataset on similar lines. India has been the largest producer of movies in the world for the last few years with a lot of diversity in languages and viewers. As per the UNESCO cinema statistics~\cite{term9}, India produces around 1,724 movies every year with as many as 1,500 movies in Indian regional languages. India's importance in the global film industry is largely because India is home to Bollywood in Mumbai. There's a huge base of audience in India with a population of 1.3 billion which is evident by the fact that there are more than two thousand multiplexes in India where over 2.2 billion movie tickets were sold in 2016. The box office revenue in India keeps on rising every year. Therefore, there is a huge need for a dataset like Movielens in Indian context that can be used for testing and bench-marking recommendation systems for Indian Viewers. As of now, no such recommendation system exists for Indian regional cinema that can tap into the rich diversity of such movies and help provide regional movie recommendations for interested audiences.
\subsection{Motivation}
As of now, Netflix and Movielens datasets do not have a comprehensive listing of regional productions as the clipping shows in Figure 1( borrowed from [39]). Therefore, a substantial source of such a data comprising movies of various regions, varying languages and genres encompassing a wider folklore is strongly needed that could provide such data in a suitable format required for building and benchmarking recommendation systems.\\
To capture the diversity of Indian regional cinema, popular websites like Netflix are trying to shift focus towards it [36, 37]. The goal is to bring some of the greatest stories from Indian regional cinema on a global platform. Through this, viewers are exposed to a wide variety of new and diverse stories from India. As a result of this initiative, Indian regional cinema will be available across countries. Building a recommendation system using a dataset of such movies and their audience can prove to be useful in such situations. Here, we present such a dataset which is the first of its kind.
\begin{figure}[!h]
\centering
        \includegraphics[width=7cm, height=9cm]{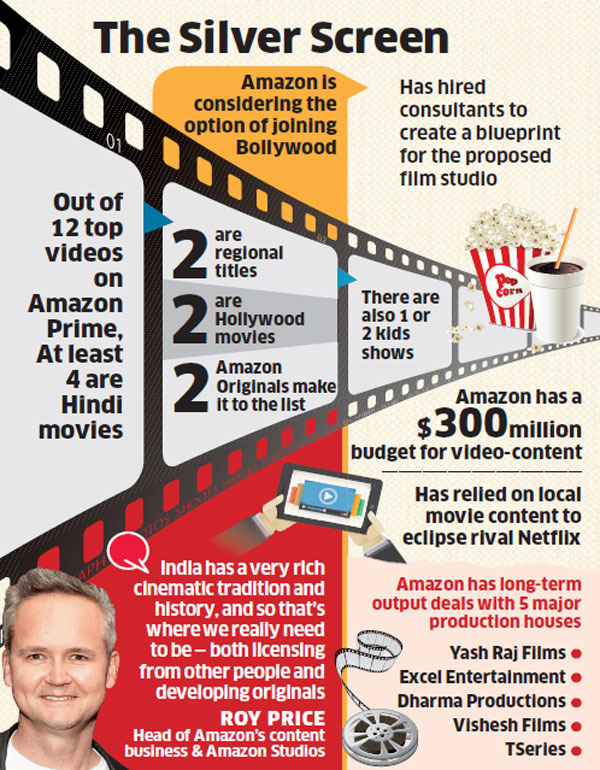}
        \caption{Amazon and Netflix: Focus on Regional Films}
\end{figure}
\subsection{Contributions}
\begin{itemize}
\item \textit{Web portal for data collection:} A web portal where a user can sign up by filling details like email, date of birth, gender, home town, languages known and occupation. The user can then provide rating to movies as like/dislike.
\item \textit{Indian Regional Cinema Dataset:} It is the first dataset of Indian Regional Cinema which contains ratings by users for different regional movies along with user and movie metadata. User metadata is collected while signing up on the portal. Movie metadata consists of genre, release year, description, language, writer, director, cast and IMDb rating.
\item Detailed analysis of the dataset using some supervised and unsupervised Collaborative Filtering techniques.
\end{itemize}
\section{Related Work}
MovieLens [1] is a web-based portal that recommends movies. It uses the film preferences of its users, collected in the form of movie ratings and preferred genres and then utilizes some collaborative filtering techniques to make movie recommendations. The Department of Computer Science and Engineering at the University of Minnesota houses a research lab known as Grouplens Research that created Movielens in 1997 [38]. Indian Regional Cinema dataset is inspired from Movielens. The primary goal was to collect data for performing research on providing personalized recommendations. MovieLens released three datasets for testing recommendation systems: 100K, 1M and 10M datasets. They have released 20M dataset as well in 2016. In the dataset, users and movies are represented with integer IDs, while ratings range from 1 to 5 at a gap of 0.5. \\
Netflix released a training data set for their contest, Netflix Prize~\cite{term8}, which consists of about 100,000,000 ratings for 17,770 movies given by 480,189 users. Each rating in the training dataset consists of four entries: user, movie, date of grade, grade. Users and movies are represented with integer IDs, while ratings range from 1 to 5.  \\
These datasets are largely for hollywood movies and TV series, and their viewers. They are not designed for those user communities which are inclined towards watching Indian regional cinema. \\
From the view point of recommender systems, there have been a lot of work using user ratings for items and metadata to predict their liking and disliking towards other items [4, 5, 6, 11]. Many unsupervised and supervised collaborative filtering techniques have been proposed and benchmarked on movielens dataset. Here, in this paper, we have chosen few popular techniques such as user-user similarity to establish baseline and then other deeper techniques such as Blind Compressed Sensing, Probabilistic Matrix Factorization, Matrix completion, Supervised Matrix Factorization are used on our dataset to provide benchmarking results. These techniques are chosen over others because these techniques have proven to provide better accuracy in recent works [6]. 
\section{Indian Regional Movie Dataset}
 This is the first dataset of Indian regional cinema which covers movies of 18 different regional languages and a variety of user ratings for such movies. It consists of 919 users with varying demographics and 2,851 movies with different genres. It has 10K ratings from 919 users.
\subsection{Metadata Information}
The data for movies has been scraped from IMDb ~\cite{term3}. IMDb has a collection of Indian movies spanning across multiple Indian regional languages and genres. Each movie is associated with the following metadata.

\begin{itemize}
    \item \textit{Movie id}:  Each movie has a unique id for its representation.
    \item \textit{Description}: Description of the movie for users.
    \item \textit{Language}: Language(s) used in the movie. A movie may have been released in multiple regional languages. The distribution is shown in Table 1.
    \item \textit{Release date}: Date of release of the movie.
    \item \textit{Rating count}: As per IMDb, to judge the popularity of the movie.
    \item \textit{Crew}: Director, writer and cast of the movie. 
    \item \textit{Genre}: Movie genre. It can be one or multiple out of 20 genres available on IMDb. The number of movies for each genre are shown in Figure 2.
\end{itemize}
\begin{figure*}[!ht]
\centering
        \includegraphics[width=15cm, height=7cm]{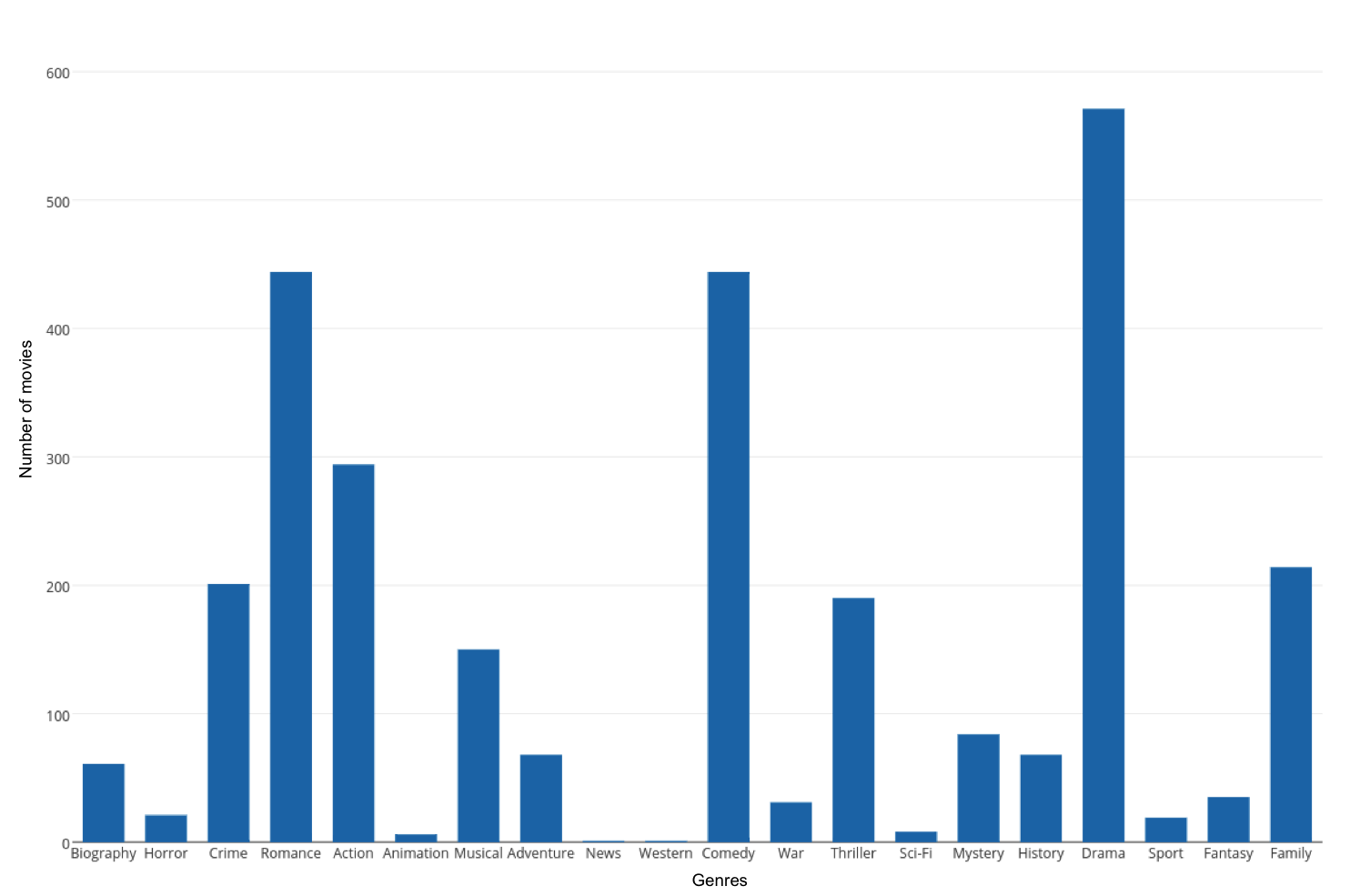}
        \caption{Genre Distribution}
\end{figure*}

\begin{table}[!h]
\caption{Language Distribution}

\begin{center}
\begin{tabular}{ |c|c|c| } 

 \hline
\textbf{Languages} & \textbf{Movie Count} & \textbf{User Count} \\
 \hline
Hindi & 615 & 902\\
\hline
Bengali & 582 & 28\\
\hline
Assamese & 22 & 9\\
\hline
Tamil & 313 & 30\\
\hline
Nepali & 51 & 9\\
\hline
Punjabi & 150 & 78\\
\hline
Rajasthani & 18 & 14\\
\hline
Malayalam & 346 & 16\\
\hline
Bhojpuri & 26 & 21\\
\hline
Kannada & 303 & 11\\
\hline
Haryanvi & 3 & 18\\
\hline
Manipuri & 8 & 4\\
\hline
Urdu & 129 & 23\\
\hline
Marathi & 204 & 14\\
\hline
Telugu & 338 & 18\\
\hline
Oriya & 98 & 6\\
\hline
Gujarati & 49 & 7\\
\hline
Konkani & 6 & 4\\
\hline
\end{tabular}
\end{center}
\end{table}

For better recommendations, it is important to include the factors which influence user ratings the most. Following is the metadata information collected for a user:
\begin{itemize}
    \item \textit{User id}: A user will have a unique id for its representation.
    \item \textit{Languages}: The languages known by the users. Its count is shown in Table 1.
    \item \textit{State}: The state of India that the user belongs to. The region wise distribution is shown in Figure 3. 
    \begin{figure}[!ht]
\centering
        \includegraphics[width=7cm, height=5cm]{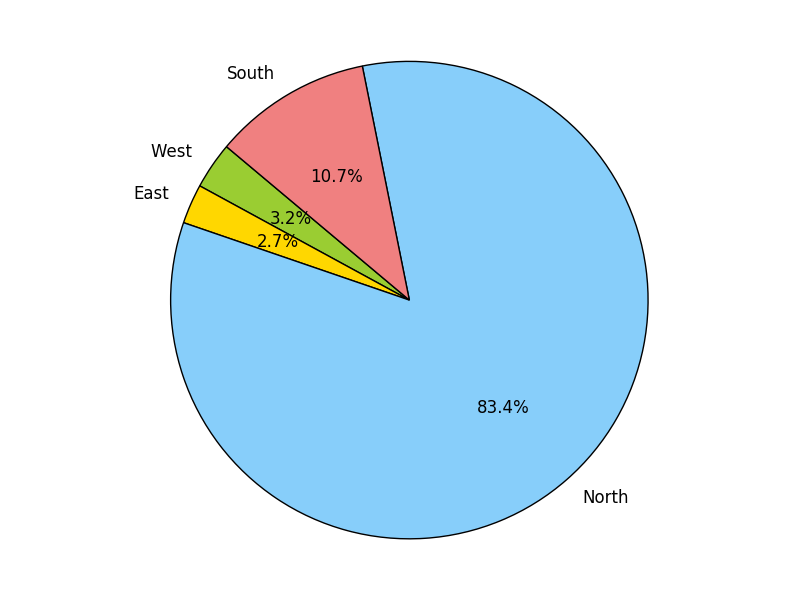}
        \caption{Region-wise Distribution}
\end{figure}
    \item \textit{Age}: Date of birth of the user is taken as input to calculate the age of the user. The distribution is shown in Figure 4.
    \begin{figure}[!ht]
\centering
        \includegraphics[width=7cm, height=5cm]{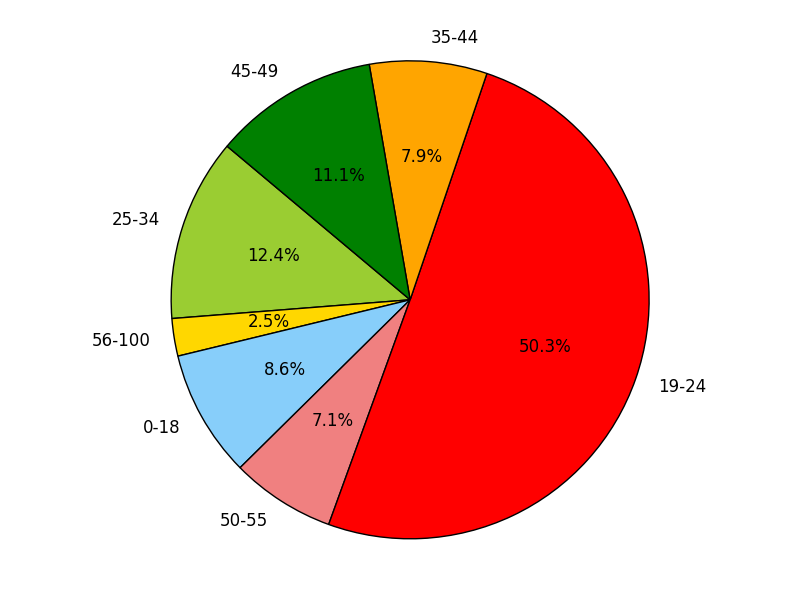}
        \caption{Age Distribution}
\end{figure}
    \item \textit{Gender}: Denotes the gender of the user. Gender distribution of the data is shown in Figure 5.
    \begin{figure}[!ht]
\centering
        \includegraphics[width=9cm, height=6cm]{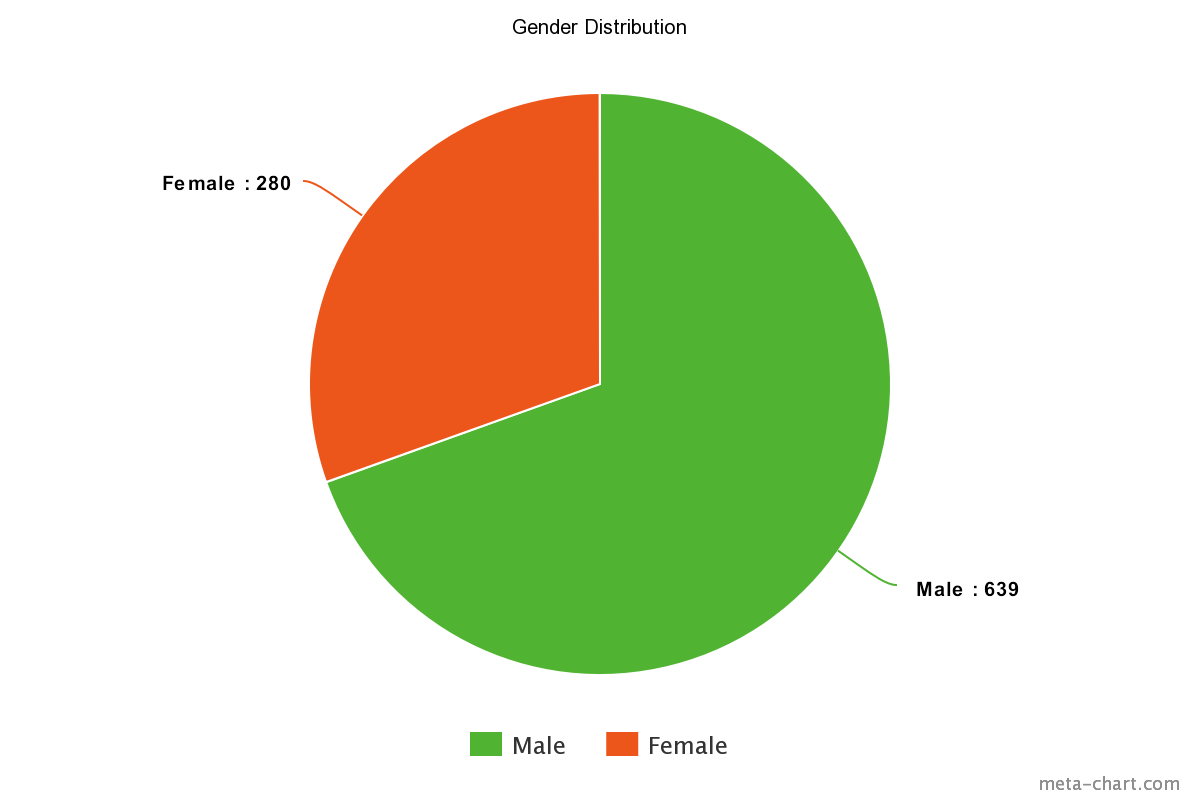}
        \caption{Gender Distribution}
\end{figure}
    \item \textit{Occupation}: It denotes the occupation of the user. It can be any one out of student, self-employed, service, retired and others. Its distribution is shown in Figure 6.
    \begin{figure}[!ht]
\centering
        \includegraphics[width=7cm, height=5cm]{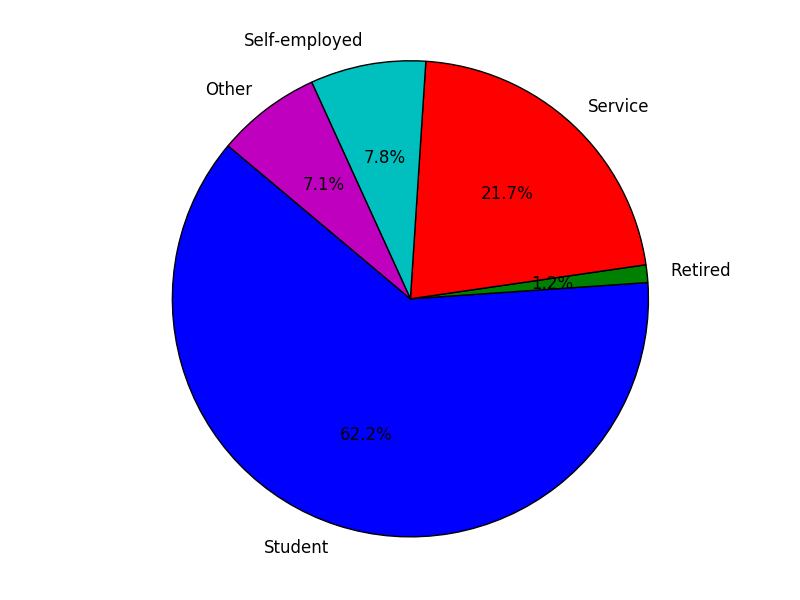}
        \caption{Occupation Distribution}
\end{figure}
\end{itemize}
\begin{table*}[!h]
\caption{Comparison of Datasets}

\begin{center}
\begin{tabular}{ |c|c|c|c|c|c| } 

 \hline
\textbf{Dataset} & \textbf{Movie Count} & \textbf{User Count} & \textbf{Rating Count} & \textbf{Sparsity(\%)} & \textbf{Release Year}\\
 \hline
Our Dataset & 2851 & 919 & 10,000 & 99.96 & 2017\\
\hline
Movielens 100K & 1700 & 1000 & 100,000 & 99.94 & 4/1998\\
\hline
Movielens 1M & 6000 & 4000 & 1,000,000 & 99.96 & 2/2003\\
\hline

\end{tabular}
\end{center}
\end{table*}
\subsection{MovieLens vs Indian Regional Cinema Dataset}
The key difference between the presented dataset and movielens is that the latter does not contain movies from Indian regional cinema. Movielens only has few Hindi and Urdu movies. Also, our data has been collected mainly from the viewers of regional movies in India. The user metadata, thus, collected can be used to recommend more relevant movies for such audiences. \\
Also, the MovieLens datasets are biased towards a certain category of users. They contain data only from users who have rated at least twenty movies. The datasets do not include the data of those users who could not find enough movies to rate or did not find the system easy enough to use. There is a possibility that there is a fundamental difference between such users and the other users in the datasets. Our dataset makes no such distinction among users based on the number of movies that they have rated. \\
To make the process of rating multiple movies easier for a user, we have used the concept of binary rating for movies where, a user can either "like" or "dislike" a movie denoted by "1" and "-1" in the dataset. On the other hand, MovieLens uses a 10-point scale for rating (from 0 to 5). A basic comparison of these datasets are shown in Table 2. The table indicates the number of users, movies, ratings, release year and the sparsity of datasets. \\

\subsection{Dataset Collection}
For the collection of user information and movie ratings, a web portal named Fickscore [15] is created where users can sign up filling in all details as shown in Figure 7. The user has to provide the preferred languages so that the portal can ask users to rate the movies of their preferred languages.\\
\begin{figure}[!ht]
\centering
        \includegraphics[width=8cm, height=8cm]{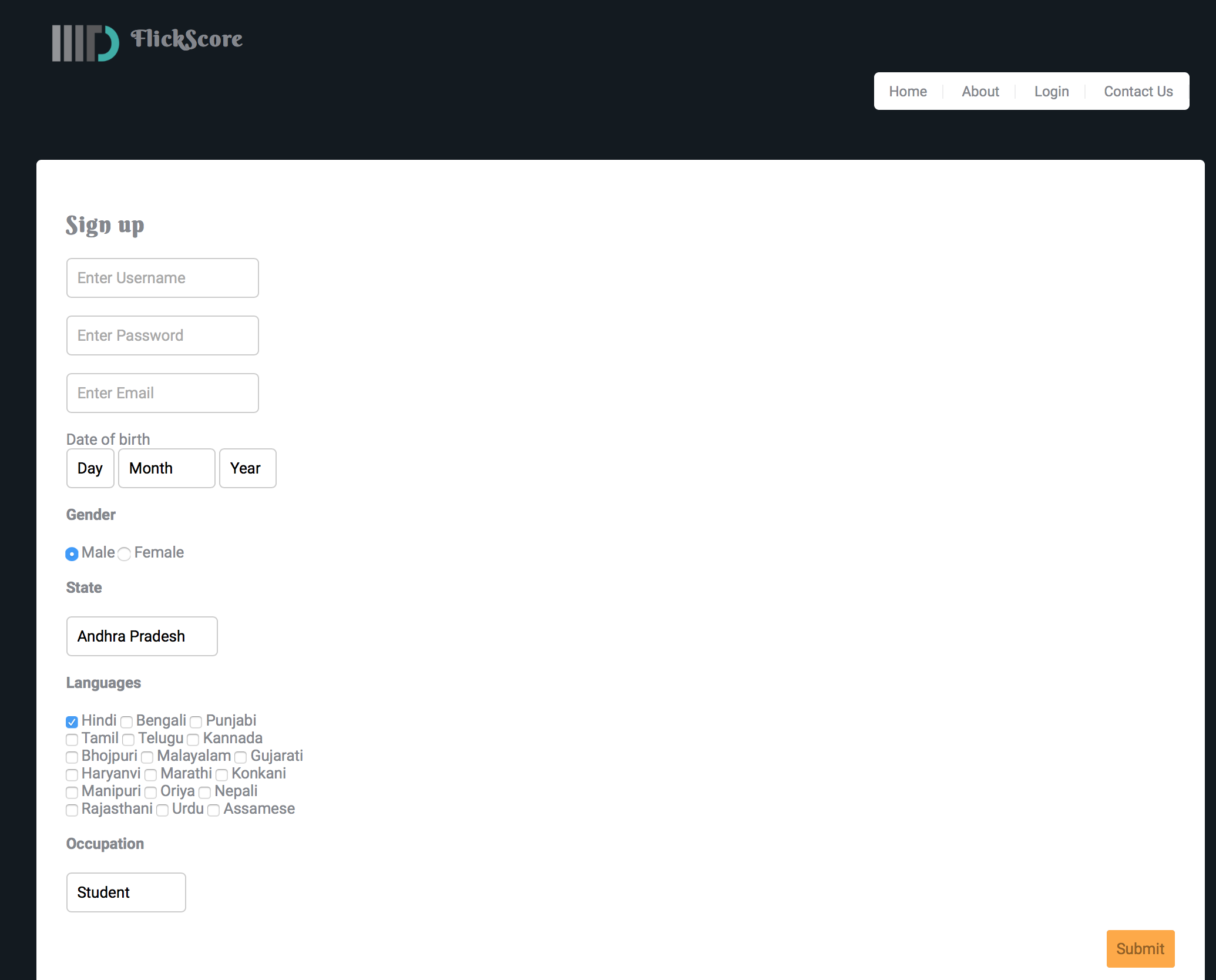}
        \caption{Sign Up form on Portal}
\end{figure}
While signing up, the user is prompted to fill up the metadata information. The user can then login to the portal to rate movies as either like or dislike and the responses are recorded as shown in Figure 8. 
\begin{figure}[!ht]
\centering
        \includegraphics[width=7cm, height=7cm]{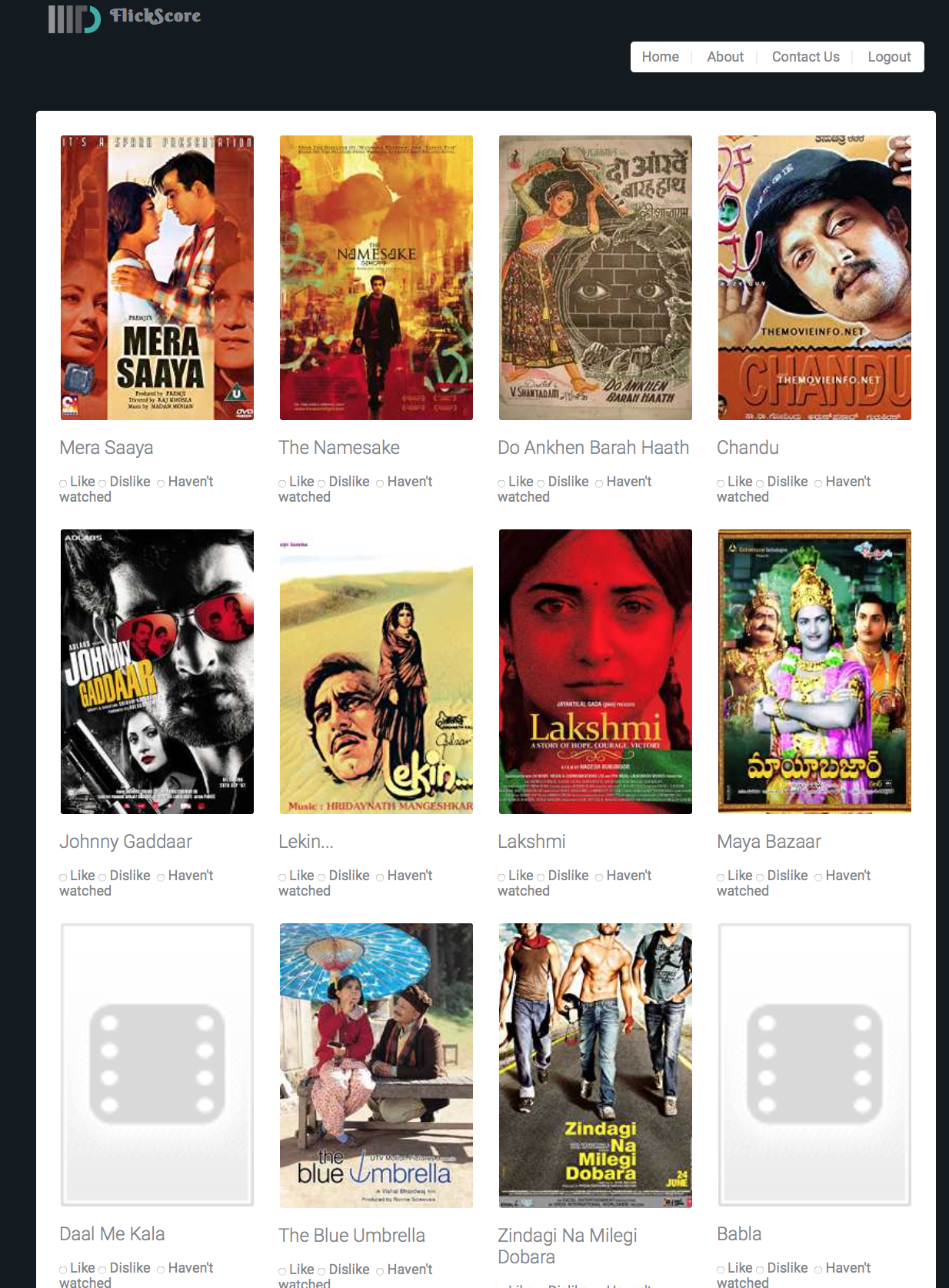}
        \caption{Rating Movies on Portal}
\end{figure}
\section{Unsupervised Collaborative Filtering techniques}
To analyze the dataset, some unsupervised techniques are used such as user-user similarity, item-item similarity, Matrix Factorization, Probabilistic Matrix Factorization, Blind Compressed Sensing etc. The main advantage of using such techniques is the ease of implementation and their incremental nature. On the other hand, it is human data dependent its performance decreases on increase of sparsity of data. These techniques cannot address the cold start problem i.e., when a new user or item adds in the dataset whose ratings are not available because these use ratings of users to make predictions. Bias correction is performed on the dataset by calculating global mean, user bias and item bias and then the above techniques are used to predict the rating of a new user for an item.

\subsection{User and Item-based similarity}
In user-user model, a similarity matrix $A$ is calculated, each entry $A_{ij}$ indicates the score computed by cosine similarity between a user $i$ and another user $j$. It denotes how much similar are two users $i$ and $j$, higher the score higher is the similarity. Similarly, in item-item model, each entry $A_{ij}$ of the similarity matrix 'A' denotes the cosine similarity score between an item $i$ and another item $j$. Higher the score, the two items are more similar.\\
Cosine similarity can be calculated for two users u and u' using the following equations:
\begin{equation}
    S_{u,u'}=  \frac{\sum_{\forall j} r_{u,j} r_{u',j}}{\sqrt {\sum_{\forall j} r_{u,j}^2 \sum_{\forall j} r_{u',j}^2}}
\end{equation}
Where, $r_{i,j}$ denotes the rating by $i^{th}$ user for $j^{th}$ item.\\
Prediction for $u^{th}$ user for $j^{th}$ item is done as:
\begin{equation}
    \hat{r_{u,j}} = \sum_{u' \in N(u), j \in R(u')} w_{u,u'}r_{u,u'}
\end{equation}

Where, $w_{u,u'}$ is the normalized similarity weight, $r_{u,u'}$ is the rating by u' user for $j^{th}$ item. 



Similar to user-user similarity, item-item similarity is calculated by computing cosine similarity between two items and ratings are predicted in the similar way.




\subsection{Matrix factorization}
There are some hidden traits (latent factors) of liking/disliking of users which may depend on the pattern of their ratings. Users and movies are mapped by this model
to a joint latent factor space. Each item $i$ and user $j$ is associated with vector $q$ and $p$ respectively which measures the possessiveness of an item or user for those factors. The dot product $q_{i}^{T} p_u $ denotes the liking of a user for a specific item which approximates the rating $r_{ui}$ ~\cite{term10}. Computing the mapping
of each user and item to factor vectors is a major challenge. Imputation
can prove to be expensive as it noticeably increases
the amount of data during calculation. To model the observed ratings directly with regularization, the following equation is used:
\begin{equation}
    \min_{q^*.p^*} \sum_{(u,i \in k)} (r_{ui}-q_{i}^{T} p_u)^2 + \lambda (|| q_i||^2 + || p_u||^2)
\end{equation}
Here, $k$ is the set of those user-item pairs in the training set for which $r_{ui}$ is known. \\
The system uses the already observed ratings to fit a model on them and uses that model to predict the new ratings. \\
The intuition behind using matrix factorization to analyze this dataset is that there should be some latent features that determine how a user rates an item. For example, two users may give high ratings to a certain movie if they both like the actors/actresses of the movie, or if the movie is an action movie, which is a genre preferred by both users. Hence, if we can discover these latent features, we should be able to predict the rating given by a certain user to a certain item, because the features associated with the user should match with the features associated with the item.\\


\begin{table*}[!h]
\centering
\caption{Unsupervised Techniques}
\begin{tabular}{|l|l|l|l|l|l|l|}
\hline
\multicolumn{1}{|r|}{\textbf{Techniques}} & \multicolumn{2}{c|}{\textbf{Movielens 100K}} & \multicolumn{2}{c|}{\textbf{Our Dataset}} & \multicolumn{2}{c|}{\textbf{Movielens 1M}} \\ \hline
      \textbf{Errors}                 & \textbf{MAE}         & \textbf{RMSE}        & \textbf{MAE}         & \textbf{RMSE}       &     \textbf{MAE}    & \textbf{RMSE}       \\ \hline
User-User similarity                   & 0.6980      & 1.026       & \textbf{0.5307}      & 1.03       & 0.607      & \textbf{0.8810}     \\
\hline
Item-Item similarity                  & 0.744       & 1.061       & \textbf{0.648}       & 1.049      & 0.671      & \textbf{0.9196}     \\ \hline
Matrix Factorization                   & 0.828       & 1.128       & \textbf{0.471}       & 0.971      & 0.6863     & \textbf{0.8790}     \\ \hline
Probabilistic Matrix Factorization                    & 0.7564      & 0.9639      & \textbf{0.481}       & 0.9372      & 0.7241     & \textbf{0.9127}     \\ \hline
Blind Compressed Sensing                    & 0.7356      & 0.9409      & \textbf{0.463}       & 0.9612     & 0.6917     & \textbf{0.8789}     \\ \hline
Matrix Completion                     & 0.8324      & 1.102       & \textbf{0.4827}      & 0.9264     & 0.7196     & \textbf{0.9102}     \\ \hline
\end{tabular}
\end{table*}

\begin{table*}[!h]
\centering
\caption{Supervised Technique}

\begin{tabular}{|l|l|l|l|l|l|l|}
\hline
\multicolumn{1}{|r|}{\textbf{Technique}} & \multicolumn{2}{c|}{\textbf{Movielens 100K}} & \multicolumn{2}{c|}{\textbf{Our dataset}} & \multicolumn{2}{c|}{\textbf{Movielens 1M}} \\ \hline
                \textbf{Errors}       & \textbf{MAE}         & \textbf{RMSE}        & \textbf{MAE}         & \textbf{RMSE}       & \textbf{MAE}        & \textbf{RMSE}       \\ \hline
Supervised Matrix Factorization          & 0.7199      & 0.9196      & \textbf{0.4367}      & 0.9283     & 0.6709     & \textbf{0.8567}     \\ \hline
\end{tabular}
\end{table*}

\subsection{Probabilistic Matrix Factorization}
It can handle large datasets because it scales linearly with the number of observations in the dataset. Let $R_{ij}$ represent
the rating of a user $i$ for a movie $j$. Let $U$ and $V$ be latent feature matrices for user and movie, respectively. The column vectors are denoted as $U_i$, representing user-specific latent feature vectors, and $V_j$, representing movie-specific latent feature vectors.~\cite{term4}. The log posterior is maximized over movie and user features with hyper parameters using the following equation:
\begin{equation}
    \begin{aligned}
\frac{1}{2} \sum_{i=1}^{N} \sum_{j=1}^{M} I_{ij} (R_{ij}-U_{i}^{T} V_j)^2 + \frac{\lambda_u}{2} \sum_{i=1}^{N} || U_i||_{frob}^2 + 
 \frac{\lambda_v}{2} \sum_{j=1}^{M} || V_j||_{frob}^2
\end{aligned}
\end{equation}
Where, $\lambda_u$ and $\lambda_v$ are the regularization parameters for user and item respectively. A local minimum of the equation can be computed by gradient descent in $U$ and $V$. The model performance is measured by computing mean average error (MAE) and root mean squared error (RMSE) on the test set. \\ This model can be viewed as a probabilistic extension of the SVD model, since if all ratings have been observed, the objective given by the equation reduces to the SVD objective in the limit of prior variances going to infinity. This technique better addresses the sparsity and scalability problems and thus improves prediction performance. It gives an intuitive rationale for recommendation.




\subsection{Blind Compressed Sensing}
A dense user item matrix is not a reasonable assumption as each user will like/dislike a trait to certain extent [24]. However, any item will possess only a few of the attributes and never all. Hence, the item matrix will ideally have a sparse structure rather than a dense one as formulated in earlier works.\\
The objective of this approach is to find the user and item latent factor matrices.
As per the approach, user latent factor matrix can be dense but the same does not logically follow for the item latent factor matrix. The sparsity of the item latent factor matrix increases the recommendation accuracy significantly. ~\cite{term5}. The following equation is minimized:
\begin{equation}
    \min_{(U,V)} ||Y-A(UV)||^2 + \lambda_u||U||_{frob} + \lambda_v||V||_{frob}
\end{equation}
Where $\lambda_u$ and $\lambda_v$ are regularization parameters for user and item respectively. $A$ is the binary mask matrix and $Y$ is the rating matrix. $U$ and $V$ is the user latent matrix and item latent matrix respectively which were assumed to be dense in earlier models. 



\subsection{Matrix Completion}
Matrix completion involves filling up the missing entries of a partially observed matrix. It aims to compute the matrix with the lowest rank or, if the rank of the completed matrix is known, a matrix of rank $r$ that matches the known entries. A popular approach for solving the problem is nuclear-norm-regularized (NN) matrix ~\cite{term7} as shown in the following equation.
\begin{equation}
    \min ||B-R||_{frob}^2 + \lambda||R||_{NN}
\end{equation}
Where, 
\begin{equation}
    B= R_{k-1} + M^T (Y- M.R_{k-1})
\end{equation}
and, $M$ is the binary mask. R is the rating matrix imputed and Y is the original rating matrix.



\section{Supervised Collaborative Filtering techniques}
To analyze the dataset, some supervised techniques are used such as supervised Matrix Factorization. The main advantage of using supervised methods is that whenever a new user or new item comes in, it can make predictions for them as well which unsupervised techniques fail to do [28, 29, 30, 31]. This is also called as cold start problem. These are scalable and are dependent on the metadata information of user and item because of which it gives more accurate predictions as it establishes relation well. Bias correction is performed on the dataset by calculating user bias and item bias and then the above technique is used to calculate the rating of a new user for an item.

\subsection{Supervised Matrix Factorization}
The task of predicting ratings becomes difficult largely because of the sparsity of the ratings available in the database of a recommender system. Therefore, using the knowledge related to users demographics and item categories can enhance prediction accuracy [25, 26, 27, 32].  Classes are formed as per users age group, gender and occupation. A user can belong to multiple classes at a time. Class label information is important to learn the latent factor vectors of users and movies in a supervised environment, in a way that they are consistent with the class label information available. Class label information puts in additional constraints which results in reducing the search space as a result of which determinacy of the problem is reduced.\\
Mathematically, within the matrix factorization framework, additional information of user metadata (U) and item metadata (V) can be used and the following equation can be minimized ~\cite{term6}.
\begin{equation}
\begin{aligned}
    \min_{(U,V,A,C)} ||Y-M(UV)||_{frob} + \lambda_u||U||_{frob} + \lambda_v||V||_{frob} +\\ \mu_u||W-UC||_{frob} + \mu_v||Q-AV||_{frob}
    \end{aligned}
\end{equation}
Where, $W_{ij}=1$ if user $i$ belongs to class $j$ else 0. $C$ is the linear map from latent factor space to classification domain. $Q$ is the class information matrix created similar to $W$. Other variables have their usual meanings. Introducing supervised learning into the latent factor model helps in improving the prediction accuracy by reducing the problem of rating matrix sparsity. The value of regularization parameters are determined using l-curve technique \cite{term16}



\section{Experiments and Results}
Three different datasets are used to compare the results of supervised and unsupervised collaborative filtering techniques used to predict user ratings. The datasets used for experiments are Movielens 100K, MovieLens 1M and our dataset of Indian regional movies. For error calculation, Mean absolute error (MAE) and Root mean squared error (RMSE) is calculated between the actual ratings and the predicted ratings. The datasets are divided into 5 folds for evaluation. The ratings are binarized into like/dislike (1/-1) labels for experiments. Results of different techniques on these datasets are shown in Table 3 and Table 4. \\ As the values in the Table 2 indicate, the basic cosine similarity measures between users and movies perform fairly well on all datasets. The minimum MAE values result from the experiments using our dataset. Since the sparsity of regional cinema dataset and Movielens 1M dataset is is very high (as indicated in Table 2), techniques like Probabilistic Matrix Factorization and Blind Compressed Sensing perform better than other basic similarity measures and among them the least MAE is again shown for our dataset. \\
To use metadata information, the information is encoded in the form of one hot vector of 1's and 0's where in case of languages, multiple 1's can be present in the vector.
Since supervised techniques uses both user and item metadata they outperform unsupervised collaborative filtering techniques. Among all three datasets, minimum MAE is shown on our dataset. This shows that the Indian Regional Cinema dataset can prove to be useful for building and benchmarking recommendation systems in Indian context, which has the most diverse languages and demographics.

\section{Conclusion and Future Work}
India is one of the country where not only varying languages are present, it's population's demographics are also very diverse in nature. Therefore, Indian regional cinema has a lot of diversity when it comes to the number of languages and the demographics of the viewers. There are thousands of such movies that are produced annually and there is a huge community of people who watch them. Therefore, a recommender system for Indian regional movies is needed to address the preferences of the growing number of their viewers. This dataset has around 10K ratings by Indian users, along with their demographic information. We believe that this dataset could be used to design, improve and benchmark recommendation systems for Indian regional cinema. We plan to release the dataset after its publication. We further want to release another version of this dataset with more number of ratings and users, which will help to improve the current state of recommender systems for the Indian audience.

\bibliographystyle{acmtrans}

\end{document}